\documentclass[twocolumn, notitlepage, floatfix, superscriptaddress, showpacs, showkeys, aps, pre, longbibliography]{revtex4-2}

\usepackage{amsmath, amssymb, bm}
\usepackage{graphicx, tabularx, multirow, makecell}
\usepackage{hyperref, cleveref}
\usepackage[table]{xcolor}
\usepackage{layouts}

\usepackage{color,soul}
\setstcolor{red}

\usepackage{kotex}

\begin{document}

\title{Gate Control Improves Routing Efficiency for Periodic Traffic}

\author{Hyunwook Song}
\thanks{These authors contributed equally to this work.}
\affiliation{CCSS, KI for Grid Modernization, Korea Institute of Energy Technology, Naju, Jeonnam 58217, Korea}

\author{Hoyun Choi}
\thanks{These authors contributed equally to this work.}
\affiliation{School of Computational Sciences, Korea Institute for Advanced Study, Seoul, 02455, Korea}

\author{B. Kahng}
\email{bkahng@kentech.ac.kr}
\affiliation{CCSS, KI for Grid Modernization, Korea Institute of Energy Technology, Naju, Jeonnam 58217, Korea}

\date{\today}

\begin{abstract}
Traffic movement on highways and data packet transmission across the Internet serve as prototypical examples of collective dynamical behavior in complex systems. Although enhancing traffic flow in complex networks has been extensively investigated, most existing studies focus on single-queue models. In this paper, we introduce a simple yet effective two-queue scheme with gate control that regulates vehicle entry and integrates it with the routing protocol. When vehicles arrive at a on-ramp gate, they first join a waiting queue and are permitted to enter the main flowing queue once it has emptied. This two-queue strategy is especially advantageous when highway inflow varies periodically with large fluctuations, as is commonly observed in real traffic. Moreover, it modifies the system’s failure characteristics, replacing sudden, discontinuous jamming transitions with gradual, continuous ones. The approach is based on a simple rule using only local queue information, making it easy to implement and widely applicable. Beyond the specific setup studied here, it can be applied to other transportation networks under different conditions or extended to other queuing systems, such as Internet data packet routing.
\end{abstract}

\maketitle

\section{Introduction}  \label{sec:info}

Complex networks with transportation phenomena are ubiquitous, ranging from biological and social systems to infrastructure. In social networks, epidemics and rumors spread through interpersonal contacts~\cite{pastor-satorras2015, moreno2004}, while metabolic pathways transport biochemical substances within cells~\cite{jeong2000}.
In engineered infrastructure, electric power grids transport electricity~\cite{watts1998a}, the Internet routes data packets~\cite{faloutsos1999a}, and highway systems facilitate vehicle movement~\cite{gastner2006}.
Consequently, optimizing traffic flow within these networks has been a subject of extensive research \cite{chen2012}. Approaches to improve traffic efficiency generally fall into two categories: modifying the network topology\cite{liu2007, zhang2007b, huang2010a, huang2010} or employing better routing strategies\cite{echenique2004, echenique2005, yan2006, danila2006, danila2007, zhang2007a, ling2010, kaitian2019}. In this study, we focus on the latter approach, as altering established infrastructure is often cost-prohibitive or impractical.

Most theoretical analyses of routing strategies are based on the packet-routing framework proposed by~\citet{ohira1998}, who identified a phase transition from a free-flow regime to a congested regime as the number of generated packets increases. Building on this model, \citet{echenique2004, echenique2005} proposed a traffic-aware routing scheme that selects paths based on neighbor congestion levels rather than on static shortest paths. This strategy was shown to delay the jamming transition and alter the nature of the phase transition.
Following this seminal work, various sophisticated routing strategies have been developed.
These include strategies based on network topology, such as efficient path routing~\cite{yan2006} and flattening betweenness centrality profiles~\cite{danila2006, danila2007}, as well as dynamic approaches that minimize estimated waiting times~\cite{zhang2007a} or total congestion~\cite{ling2010, kaitian2019}.
However, these advancements often come at the cost of increased complexity, requiring non-local or global information to optimize traffic distribution. This reliance on extensive information exchange poses significant scalability challenges in large-scale distributed systems.

Concurrently, there has been growing research interest in capturing the realistic nature of transportation dynamics, including time dependence. Unlike the constant traffic assumed in earlier models, real-world traffic exhibits strong temporal fluctuations, often driven by daily or periodic cycles~\cite{takayasu2000, demenezes2004, meloni2008}. While recent studies have attempted to address this by modeling time-variant traffic~\cite{tang2009}, they typically inherit the limitation of requiring additional information, such as adaptive routing based on the global congestion level, to be effective in the face of these fluctuations.

Previous studies in the field of traffic engineering provide a useful precedent for addressing this challenge. In highway traffic control, ramp metering regulates inflow at on-ramps and can improve traffic capacity during high demand periods by restricting vehicle entry~\cite{papageorgiou2002}. Moreover, ramp meters can be operated through local feedback control with only on local information~\cite{papageorgiou}. These studies suggest that entry regulation, rather than route optimization alone, can provide an effective mechanism for stabilizing transportation systems under time-dependent demand.

Motivated by this inflow-control principle, we propose a minimal and physically interpretable queueing model within the traffic-aware routing framework. Instead of using a single queue for both vehicle creation and transport as in prior models, we introduce a gate protocol that separates these two roles. Newly created vehicles are first held in a \textit{waiting} queue, and the gate controls their entry into the \textit{flowing} queue, where they are transported to neighboring nodes. Unlike previous methods that optimize vehicle routing paths, our gate-based approach regulates entry timing without computational overhead. This simple construction allows us to examine how local inflow control modifies jamming dynamics under periodic traffic demand while keeping the routing protocol itself unchanged.

We demonstrate that this local restriction acts as a self-regulating buffer, effectively decoupling the internal relaxation dynamics from the time-varying external demand. We find that the advantages of the proposed method increase as vehicle transportation becomes more complex. It outperforms the unrestricted case by significantly delaying the onset of congestion and lessening its overall intensity. Furthermore, it transforms the system's breakdown from a discontinuous, catastrophic collapse into a continuous, manageable accumulation. Our findings suggest that in time-varying environments, regulating entry at the microscopic level is a robust strategy that yields high efficiency with minimal information.

The rest of this paper is structured as follows: Sec.~\ref{sec:methods} describes the network model, time-dependent vehicle creation, and the specific implementations of two routing protocols (with and without gate control). Sec.~\ref{sec:macroscopic} reports macroscopic results: the jamming transitions and vehicle densities for the two protocols as a function of model parameters. Sec.~\ref{sec:microscopic} explores the microscopic processes underlying these macroscopic behaviors, with particular attention to queue dynamics and passage time distributions. Finally, Sec.~\ref{sec:discussion} recaps our main results and examines their implications for designing resilient transportation networks.

\section{Two-queue model with Periodic input}   \label{sec:methods}

\subsection{Underlying traffic Network} \label{subsec:network}
We adopt one of the simplest yet most representative Erdős–Rényi (ER) random graphs~\cite{erdos2011} as the underlying structure of the traffic network. Each node represents an interchange or highway junction with gates, and vehicles travel along roads (links) that connect these nodes. To avoid technical complications in the analysis, we remove leaf nodes with a single neighbor using $k$-core pruning~\cite{seidman1983a} with $k=2$.
We confirm that our core findings are robust across other topologies, such as scale-free networks, as detailed in Appendix~\ref{app:settings}.

\subsection{Vehicle creation}  \label{subsec:vehicle}
In this study, we incorporate time-varying vehicle entrances to capture traffic fluctuations driven by daily cycles~\cite{tang2009}.
Specifically, the number of vehicles created in the system at each time step $t$ is given by
\begin{equation}    \label{eq:periodic-p}
    p(t) = p_0\left[1+\eta\sin{\left(\frac{2\pi t}{T}\right)}\right],
\end{equation}
where $p_0$ is the average number of vehicles created per step, $0 \le \eta \le 1$ is the amplitude, and $T$ is the external driving period measured in simulation steps.
When $\eta=0$, the vehicle generation rate is constant $p_0$.
Varying $T$ leaves $p_0$ and the routing and gate update rules at each step unchanged.
For a daily input cycle, larger $T$ therefore corresponds to a higher average vehicle generation rate and more frequent network updates in physical time.
We consider $T=24$, $288$, and $1440$, corresponding to 1 hour, 5 minutes, and 1 minute per network update, respectively.

\subsection{Routing protocols}  \label{subsec:protocols}

We use two types of queues at each node: a \textit{waiting} queue, which holds vehicles before they can pass through the gate, and a \textit{flowing} queue, which transports vehicles from one node to another. Both queues operate on a first-in, first-out discipline. The number of vehicles in the waiting and flowing queues of each node $i$ is denoted by $w_i$ and $f_i$, respectively. At each time step $t$, $p(t)$ vehicles are created, and each of them is assigned to a distinct, randomly selected node. Its destination is assigned at random, except for its origin node.

\begin{figure}
    \centering
    \includegraphics[width=0.7\linewidth]{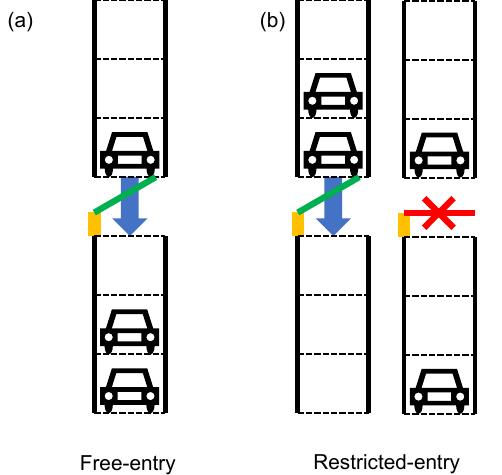}
    \caption{
    Two gate control protocols. (a) In the free-entry (FE) protocol, the gate remains open regardless of the flowing queue state, enabling unrestricted entry from the waiting queue. (b) In contrast, the restricted-entry (RE) protocol permits entry to the flowing queue only when it is empty ($f_i = 0$).
    }
    \label{fig:schematic}
\end{figure}

We consider two protocols that differ in their gate control as schematically depicted in Fig.\ref{fig:schematic}. In the free-entry (FE) protocol, the gate is always open, thereby recovering the conventional single-queue model. On the other hand, in the restricted-entry (RE) protocol, the gate permits entry to the flowing queue only when it is empty ($f_i = 0$).

Once vehicles enter the flowing queue, their transportation follows the traffic-aware routing of~\citet{echenique2005}. Each vehicle with destination $j$ advances to a neighboring node $i$ that minimizes the effective distance $\delta_{ij}$, defined as
\begin{equation}    \label{eq:effective_distance}
    \delta_{ij} \equiv hd_{ij} + (1-h)f_i.
\end{equation}
$d_{ij}$ represents the length of the shortest path between nodes $i$ and $j$.

The tunable parameter $0 \le h \le 1$ quantifies the extent of traffic awareness.
For $h=1$, vehicles follow the shortest path regardless of congestion. In contrast, when $h<1$, a large flowing queue at node $i$ deters incoming vehicles, allowing them to choose alternative paths if that node is congested. It is well established that incorporating traffic awareness ($h<1$) results in more efficient routing. When a vehicle arrives at its destination, it immediately disappears. In this study, we assess and contrast the traffic efficiency of the RE and FE for a fixed value of $h=0.8$, as a representative value.
We also verify the robustness of our results under a different traffic-awareness parameter, $h=0.5$, in Appendix~\ref{app:settings}.

\section{Macroscopic Behaviors} \label{sec:macroscopic}

\begin{figure}
    \centering
    \includegraphics[width=\linewidth]{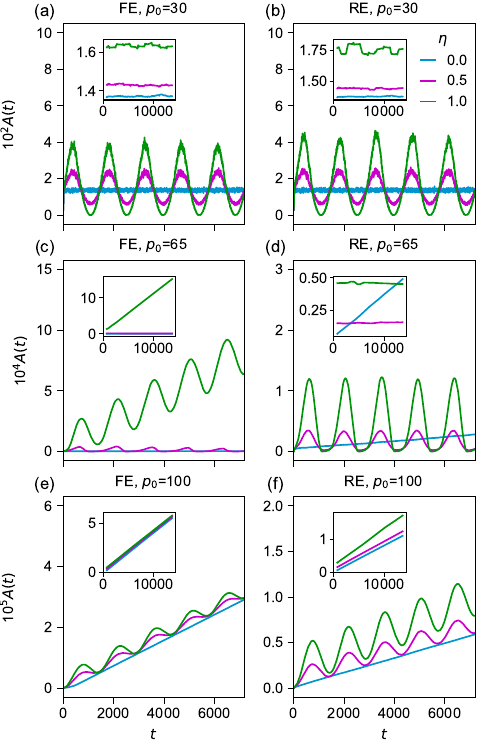}
    \caption{
Temporal evolution of the total number of vehicles in queues, $A(t)$, as a function of time step $t$ with a driving period of $T = 1440$.
Results for the free-entry (FE) protocol are shown in left column and for the restricted-entry (RE) protocol in right column. Each color represents a different amplitude $\eta$ (see legend in (b)). Insets show the time average of $A(t)$ with window length $T$. At intermediate $p_0$ ((c), (d)), FE and RE exhibit opposite trends with increasing $\eta$ due to the asymmetric behavior in Fig.~\ref{fig:order-parameter}(e) and (f).
} \label{fig:dynamics}
\end{figure}

The number of vehicles in the flowing and waiting queues is denoted by
\begin{equation}
    F(t) \equiv \sum_{i=1}^{N}{f_i(t)}, \ W(t) \equiv \sum_{i=1}^{N}{w_i(t)},
\end{equation}
and the total number of vehicles in the system $A(t) \equiv F(t) + W(t)$.

\subsection{Evolution of system}   \label{subsec:dynamics}

Fig.~\ref{fig:dynamics} demonstrates how $A(t)$ evolves following the two protocols, FE and RE, for different values of $p_0$ and $\eta$. $A(t)$ oscillates according to the daily modulation of $p(t)$ with $T=1440$ time steps per day. The inset in each panel presents the time-averaged $A(t)$ in units of $T$ time steps, where the linear behavior indicates that the modulation follows the daily cycle of $p(t)$. In each case, the system first passes through a transient phase and then enters a steady state in which its evolution settles into linear behavior.

For any value of $\eta$, when $p_0$ is small (Figs.~\ref{fig:dynamics}(a) and (b)), the averages of $A(t)$ for both protocols remain constant with minor fluctuations because the routing system can handle all vehicles. For large $p_0$ (Figs.~\ref{fig:dynamics}(e) and (f)), the averages of $A(t)$ for both protocols grow linearly on average while following the same daily modulation.
Remarkably, at intermediate $p_0$ (Figs.~\ref{fig:dynamics}(c) and (d)), the two protocols show opposite behavior with increasing $\eta$: while FE transitions toward congestion, RE transitions from congestion toward free-flow due to the crossover behavior in Fig.~\ref{fig:order-parameter}.

\subsection{Jamming transition} \label{subsec:phases}

\begin{figure}
\centering
\includegraphics[width=\linewidth]{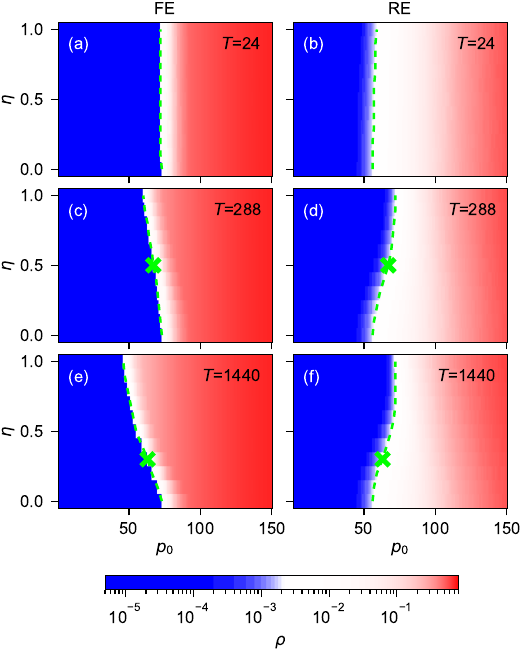}
\caption{
    Order parameter $\rho$ vs. $p_0$, $\eta$, and $T$. Green dashed curves represent the transition curves $p_c(\eta)$ for each $T$. For the FE, the order parameter increases drastically beyond $p_c$ as $p_0$ increases. This indicates that the transition is discontinuous. On the other hand, for the RE, the order parameter increases gradually beyond $p_c$ as $p_0$ increases, indicating a continuous transition. The green cross represents the point where $p_c$ of FE and RE are same.
}   \label{fig:order-parameter}
\end{figure}

The performance of each routing system is quantified by the following measure~\cite{echenique2005}:
\begin{equation}    \label{eq:order_parameter}
\rho \equiv \lim_{t \to \infty}{\frac{A(t+\tau)-A(t)}{p_0\tau}},
\end{equation}
where $\tau$ is an integer multiple of the driving period $T$.
$\rho$ denotes the increase rate of congestion, normalized by $p_0$, in a steady state.
When $\rho=0$, the system is in a free-flow state where all created vehicles arrive at their destinations without congestion. When $\rho=1$, the system is in a fully jammed state, with no vehicles arriving at their destinations. $\langle \rho \rangle$ is the ensemble average of $\rho$, which serves as the order parameter in the jamming transition. Hereafter, the bracket is eliminated for simplicity.

The transition point $p_c$ of the traffic routing system is defined as the smallest $p_0$ from which $\rho$ becomes nonzero. So, a larger $p_c$ indicates better performance.

Figs.~\ref{fig:order-parameter} are the phase diagrams of the jamming transition in the parameter space of ($p_0$, $\eta$) for different $T$ for the FE and RE protocols. The transition curve $p_c(\eta)$ is denoted by green dashed curves, and green crosses indicate the points where the $p_c$ of FE and RE are the same.

One notable difference between the two protocols is the shape of $p_c$ as a function of $\eta$. When $\eta$ is small, the transition point $p_c(\eta)$ is large for the FE, whereas it is small for the RE. Thus, the FE protocol is more effective. On the other hand, when $\eta$ is large, the RE is more effective. As $T$ increases, this tendency of $p_c$ becomes more pronounced.
We also recognize that the type of transitions shown in Fig.~\ref{fig:order-parameter} differ. For the FE, $\rho$ rises abruptly, and the transition is discontinuous. By contrast, for the RE protocol, $\rho$ increases gradually, and the transition is continuous.

\begin{figure}
\centering
\includegraphics[width=\linewidth]{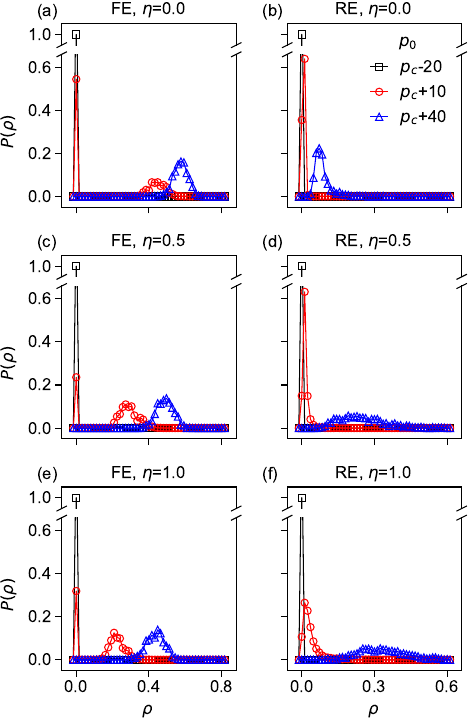}
\caption{
The order parameter distribution $P(\rho)$ at $T=1440$ for the free-entry (FE, left column) and restricted-entry (RE, right column) protocols at different amplitude values $\eta$. The FE protocol displays two distinct peaks, whereas the RE protocol exhibits a single peak.
}
\label{fig:rho-dists}
\end{figure}

We consider the distribution of the order parameter $P(\rho)$ in Fig.~\ref{fig:rho-dists}. The data are accumulated over different configurations for given $p_0$ values. For the FE, $P(\rho)$ displays two peaks at $\rho=0$ and a finite $\rho$ for all $\eta$ cases (see Fig.~\ref{fig:rho-dists} (a), (c), and (e)). The separation of the two peaks indicates that the jamming transition is discontinuous. On the other hand, for the RE, $P(\rho)$ displays a single peak for all $\eta$ cases (see Fig.~\ref{fig:rho-dists}(b), (d), and (f)).
The peak position of $\rho$ increases continuously as $p_0$ increases, exhibiting a continuous transition. This behavior is consistent with that predicted in Fig.~\ref{fig:order-parameter}.

\section{Microscopic Mechanisms}   \label{sec:microscopic}
We analyze the microscopic dynamics of the FE and RE protocols to uncover the mechanisms that give rise to the observed differences in macroscopic quantities.

\subsection{Constant vehicle creation} \label{subsec:constant-rate}

For $p(t)=p_0$, that is, when $\eta=0$, the RE protocol undergoes a transition at a lower value of $p_0$ than the FE protocol, and the natures of these transitions are distinct in the two protocols. To investigate the mechanisms underlying these differences, we introduce the following two quantities.

First, we express the increasing rate of the total queue length of node $i$ in a single time step during a steady state as
\begin{equation}
\Delta a_i \equiv \lim_{t\rightarrow\infty}{\frac{\left[f_i(t+\tau)-f_i(t)\right]+\left[w_i(t+\tau)-w_i(t)\right]}{\tau}}.
\end{equation}
The order parameter is directly associated with $\Delta a_i$ as $\rho = 1/p_0 \sum_i \Delta a_i$.

\begin{figure}
\centering
\includegraphics[width=\linewidth]{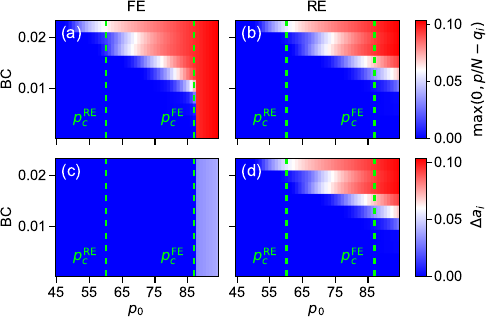}
\caption{
Microscopic behavior of FE and RE protocols at constant creation rate in a single network.
(a) and (b) Expected value of node-wise waiting queue length increment, $\max (0, p_0/N - q_i)$.
(c) and (d) Node-wise queue length increment $\Delta a_i$.
The measurements for each node are averaged based on its betweenness centrality (BC).
The transition points $p_c^\text{FE}$ and $p_c^\text{RE}$ are highlighted as a green dashed line in each plot.
} \label{fig:constant-p_0-qi_ai}
\end{figure}

Second, $q_i$ denotes the probability that the flowing queue at node $i$ is empty. In the RE protocol, this corresponds to the probability that a vehicle in the waiting queue passes through the gate into the flowing queue. By definition, the probability that a new vehicle enters the waiting queue at each time step is $p_0/N$. Consequently, $\max(0, p_0/N - q_i)$ in the RE protocol is the net probability that the length of the waiting queue at node $i$ increases.

In the previous study \cite{echenique2005}, which analyzed the FE protocol in the special case $\eta=0$, congestion was observed to arise at nodes with high betweenness centrality (BC)~\cite{freeman1977}. Here, we examine whether an analogous mechanism is responsible for congestion in our model. Specifically, we explore how nodes with large BC relate to the two quantities defined above.

\begin{figure}
    \centering
    \includegraphics[width=0.667\linewidth]{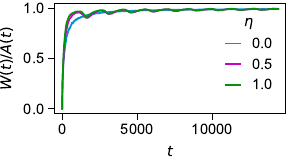}
    \caption{
        Fraction of vehicles in waiting queues throughout the dynamics of a RE system in the jammed phase for different $\eta$. Most vehicles accumulate in the waiting queues, resulting in $W/A \approx 1$.
    }   \label{fig:RE-jamming-position}
\end{figure}

The coherence is calculated by averaging the data within each unit box in the space of BC and $p_0$ (Fig.~\ref{fig:constant-p_0-qi_ai}). It indicates that $\max(0, p_0/N - q_i)$ for the FE and RE protocols are similar to one another, as long as $p_0 < p_c^\text{FE}$. This result suggests that nodes with high BC are more likely to be congested. On the other hand, the coherence between the BC and $\Delta a_i$ in the FE and RE protocols is somewhat different, as shown in Fig.~\ref{fig:constant-p_0-qi_ai}(c) and (d).
In the FE protocol, vehicles that appear in a non-empty flowing queue are allowed to enter the transfer process. Consequently, a small number of such queues does not lead to congestion. Instead, the vehicles generated more than the threshold, causing discontinuous jamming. In contrast, in the RE protocol, congestion emerges in the waiting queues at nodes with high BC, as illustrated in Fig.~\ref{fig:RE-jamming-position}. In the steady state, the proportion of vehicles residing in the waiting queues approaches $1$ in the congested state, regardless of the creation amplitude $\eta$.

Thus, $\Delta a_i$ in the RE protocol can be expressed as
\begin{equation} \label{eq:wait-jam}
\Delta a_i \approx \lim_{t \rightarrow \infty} {\frac{w_i(t+\tau)-w_i(t)}{\tau}} = \max\left(0, \frac{p_0}{N}-q_i \right),
\end{equation}
which is numerically validated in Fig.~\ref{fig:constant-p_0-qi_ai}(b) and (d).
The presence of a single node with $q_i < p_0/N$ results in congestion in that node's waiting queue and initiates the jamming transition, leading to $p_c^\text{RE}<p_c^\text{FE}$.
This simultaneously explains the continuous phase transition in the RE protocol: $\Delta a_i$ departs from zero in only a few high BC nodes, gradually increasing the global order parameter metric $\rho$.

\subsection{Periodic vehicle creation}  \label{subsec:pc_eta}

\subsubsection{Restricted Entry protocol} \label{subsec:re-pc-change}

As outlined in the above section and illustrated in Fig.~\ref{fig:RE-jamming-position}, the jamming transition of the RE protocol is induced by the increase in the waiting queue $p_0/N$ and the incoming rate $q_i$ at nodes with high BC. In the case $\eta > 0$, the incoming rate $q_i$ becomes time-dependent. Thus, Eq.~\eqref{eq:wait-jam} is obtained again when the time interval $\tau$ is taken as an integer multiple of $T$.
\begin{equation}
\Delta a_i
\approx \lim_{t \rightarrow \infty} {\frac{1}{\tau}  \sum_{t'=t+1}^{t+\tau}{
\left[\frac{p(t')}{N}-q_i(t')\right]}
}=\frac{p_0}{N}-q_i
\end{equation}
when $p_0/N>q_i$, and $\Delta a_i = 0$ otherwise.
Here, $q_i$ is re-defined as a time average of $q_i(t)$ for $\tau$ time steps in steady state.

\begin{figure}
\centering
\includegraphics[width=\linewidth]{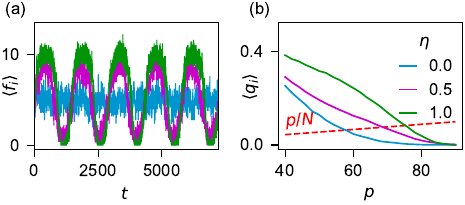}
\caption{(a) Flowing queue length $\langle f_i \rangle$ as a function of time step $t$ for $p_0=66$, and (b) probability $\langle q_i \rangle$ that the flowing queue is empty as a function of the average vehicle creation rate $p_0$. They are averaged over the five nodes with the largest BC. Different colors indicate different creation amplitudes $\eta$, from 0.0 (blue) to 1.0 (green). The red dashed line shows, for comparison, the growth of the waiting queue $p_0/N$.
}
\label{fig:RE-fluctuation}
\end{figure}

Fig.~\ref{fig:RE-fluctuation}(a) shows the temporal average of $f_i(t)$ for the five nodes with the largest BC values across different $\eta$. As the level of fluctuation grows, the time intervals during which $f_i(t) = 0$ become longer, enabling vehicles from the waiting queue to move into the flowing queue more easily. This trend is likewise observed in Fig.~\ref{fig:RE-fluctuation}(b). For a fixed $p_0$, the value of $q_i$ for the top-5 nodes increases with $\eta$, thereby postponing the transition point $p_c$ at which $p_0/N$ surpasses $q_i$.

\subsubsection{Free Entry protocol} \label{subsec:fe-pc-change}

\begin{figure}
\centering
\includegraphics[width=\linewidth]{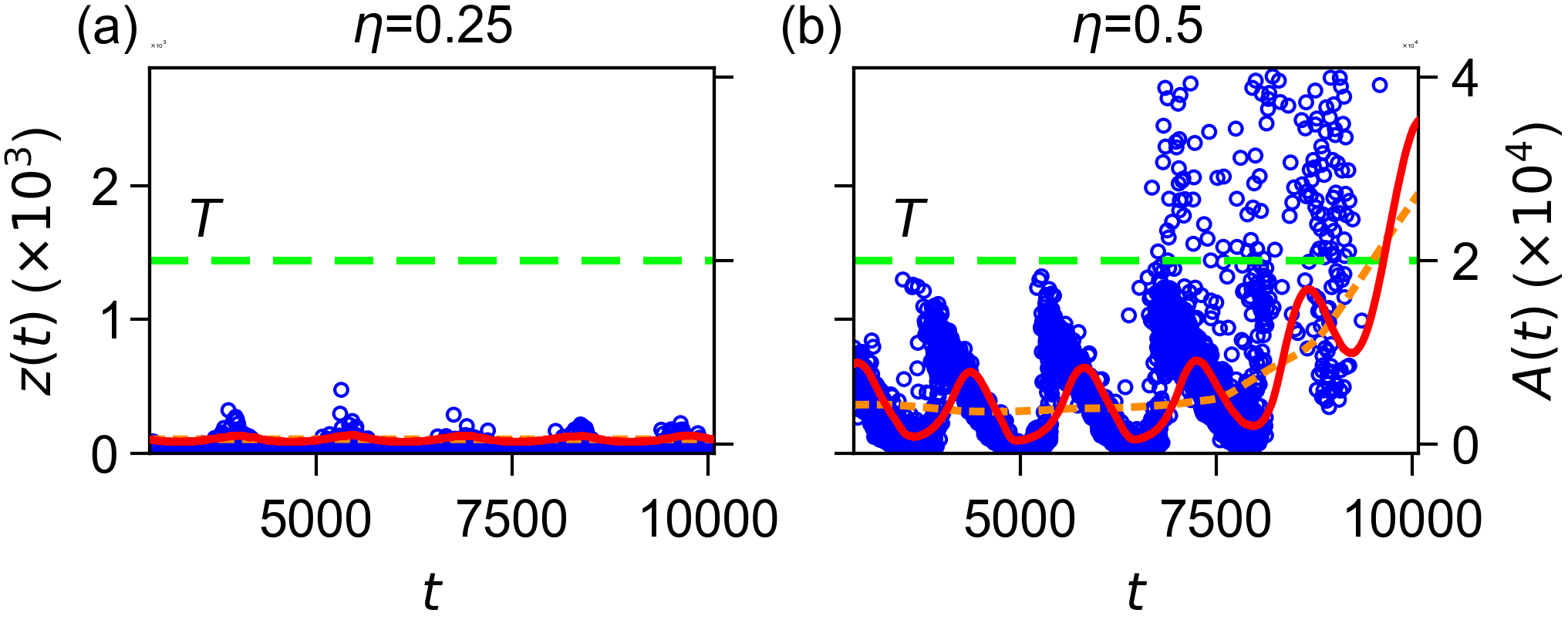}
\caption{
The longest passage time $z(t)$ of vehicles created at each time step $t$ and the number of vehicles in the system $A(t)$. (a) is at the transition point ($p_c=67$) of a system with $\eta=0.25$, and (b) is the same for $\eta=0.5$. $z(t)$ is represented by the blue scatter plot following the left $y$-axes, and $A(t)$ is represented by the red curve, following the right $y$-axes. The cyan dashed line is the threshold $T$ for the maximum passage time, and the orange line is the moving average of $A(t)$ over $T$ time steps.
}
\label{fig:FE-packet-passage}
\end{figure}

We investigate the vehicle passage time in the FE rule for $\eta > 0$, defined as the number of time steps between a vehicle’s creation and its arrival.
Let $z(t)$ denote the maximum passage time among all vehicles generated at time $t$. Fig.~\ref{fig:FE-packet-passage} shows how $z(t)$ evolves over time for different values of $\eta$ in a single realization under the FE rule.
In Fig.~\ref{fig:FE-packet-passage}(a), where $\eta$ is small, $z(t)$ only fluctuates slightly and never exceeds $T$ (indicated by the dashed line). In contrast, in Fig.~\ref{fig:FE-packet-passage}(b) with larger $\eta$, $z(t)$ occasionally approaches the congestion regime. Once $z(t)$ becomes larger than $T$ (which corresponds to one full day), the number of vehicles in the system, $A(t)$ (red curve), starts to grow. The reason is that any vehicle that fails to arrive within $T$ remains in the system, thereby blocking the routing of subsequently generated vehicles. As a result, for given $p_0$ and $T$, there is a characteristic value $\eta_c$ above which the system enters a congested state. This mechanism explains why the transition point $p_c$ decreases as $\eta$ increases, in contrast to the behavior found in the previous section for the RE systems.

\subsubsection{Effect of the driving period}

Fig.~\ref{fig:order-parameter} further demonstrates that the dependence of $p_c$ on $\eta$ becomes more pronounced as the driving period $T$ increases in both the RE and FE systems.
As discussed in previous sections, the transition points of the RE and FE systems are determined by length fluctuations in the flowing queue and the longest passage time, respectively. Both of these quantities are controlled by the cumulative value of $p(t)$,
\begin{equation}  \label{eq:pt-integral}
\begin{split}
    \sum_{t'=1}^{t} {p(t')} &\approx \int_{0}^{t} {p_0\left[1+\eta\sin{\left(\frac{2\pi t'}{T}\right)}\right]dt'} \\
    &= p_0t + p_0\frac{\eta T}{2\pi} \left[ 1 - \cos\left( \frac{2\pi t}{T} \right) \right].
\end{split}
\end{equation}
Here, the first term reflects the contribution from the average creation rate, whereas the second term, originating from fluctuations, scales with both $\eta$ and $T$. Consequently, the influence of $T$ on the flowing queue length and the passage time mirrors that of $\eta$, enhancing the variation in $p_c$ in both systems.

\section{Conclusion}    \label{sec:discussion}

In this work, we develop a traffic routing framework that explicitly differentiates between vehicles waiting to enter the network and those already moving within it by assigning each node two distinct queues: a \textit{waiting} queue and a \textit{flowing} queue.
The waiting queue stores newly created vehicles until they pass through the gate into the flowing queue, which transports vehicles to neighboring nodes.
Building on this setup, we examine two protocols under a fixed routing scheme: the Free Entry (FE) rule, which injects vehicles continuously regardless of the node’s state, and the Restricted Entry (RE) rule, which allows injection only when the flowing queue is empty. This restriction serves as a purely local control mechanism and does not rely on any global information~\cite{tang2009}.

On a macroscopic scale, the RE protocol responds to fluctuations in the external drive in a different manner compared to the FE protocol. As the volatility ($\eta$ or $T$) increases, the FE system becomes progressively more congested, and the jamming threshold $p_c$ becomes smaller.
By contrast, the RE system grows more resilient to such fluctuations, with $p_c$ delayed to higher values.
Furthermore, even after the system has entered the congested phase, the accumulation rate $\rho$ stays significantly lower under RE than under FE. Taken together, these results show that the local entry restriction functions as an effective buffering mechanism, delaying the onset of jamming and mitigating congestion severity when demand exhibits strong fluctuations close to the edge of traffic congestion.

The divergence in macroscopic behavior stems from distinct microscopic jamming mechanisms. In the constant-creation regime ($\eta = 0$), we confirmed that the transition in RE systems is initiated by a mismatch between the injection rate $p_0/N$ and the effective entry probability $q_i$ at nodes with large betweenness centrality (BC). This leads to congestion that is mostly confined to waiting queues, typically causing the transition to occur earlier than in FE systems, but with a less severe buildup of vehicles. When the drive fluctuates ($\eta > 0$), the RE system benefits from extended intervals during which traversing queues are empty, which effectively raises $q_i$ at high-BC nodes. In contrast, jamming in the FE system arises from the competition between the vehicle traversal times and the daily external driving cycle ($T$ time steps). As the fluctuations in $p(t)$ grow, the variance of traversal times increases, rendering the system more susceptible to jamming. This mechanistic distinction accounts for the crossover in $p_c$ between the two protocols as volatility grows.

Unlike previous approaches that attempt to improve routing by globally reweighting paths \cite{yan2006, danila2006, danila2007} or by employing traffic-aware cost functions \cite{zhang2007a, ling2010, kaitian2019}, our method modifies only the entry condition while keeping the routing rule intact. While the notion of restricting inflow is reminiscent of ramp metering in transportation engineering, the RE rule can be interpreted as an idealized local counterpart of such inflow-control strategies for periodically driven network traffic. We analyze this local control mechanism through the lens of non-equilibrium phase transitions in complex networks. Our findings demonstrate that a simple local constraint can qualitatively alter the phase diagram ($p_c$ vs. $\eta$) as well as the nature of the phase transition (continuous vs. discontinuous), thereby offering a new perspective on controlling congestion dynamics in network-based transport systems without changing either the underlying network topology or the routing protocols. In real transportation systems, the strict rule could be implemented directly in strongly controlled entry settings, approximated by smoother release rules such as the intermediate protocol in Appendix~\ref{app:intermediate}, or realized indirectly through dynamic access pricing based on the current flowing-queue length.

While our study demonstrates the effectiveness of restricted entry on ER graphs for particular routing parameters, confirming how broadly these results apply remains an important avenue for future work. A natural follow-up is to investigate whether the relative benefit of the RE protocol under high volatility continues to hold across a variety of network topologies—such as scale-free, small-world, or real-world transportation networks—and under more varied conditions, including heterogeneous link capacities or alternative settings of the traffic-awareness parameter ($h$).

Furthermore, extending the driving mechanism beyond global sinusoidal modulation would allow for a broader exploration of non-equilibrium network control.
Examining other temporal variations, such as stochastic noise or bursty demands, would bring to light the competition between external timescales and internal relaxation dynamics.
In addition, the spatial variation of traffic, such as that found on the Internet spanning different time zones, can be considered to study the effect of asynchronism.

In summary, our study provides systematic evidence for the seemingly paradoxical idea that imposing restrictions can improve routing performance. Although limiting vehicle entry appears at odds with the objective of maximizing throughput, we show that such localized constraints are essential for preserving stability under time-varying driving conditions. By functioning as a self-regulating buffer during periods of peak demand, the restricted-entry mechanism not only postpones the onset of traffic jams but also reduces the intensity of congestion once the critical threshold is exceeded. Our results thus establish that controlling access to the active transport layer is as influential as optimizing routing paths, and together they offer a robust framework for handling non-stationary traffic dynamics.


\begin{acknowledgments}
This work was supported by the National Research Foundation of Korea (NRF) funded by the Korea Government (MSIT) grant No. RS-2023-00279802 and No. RS-2026-25556154 (BK) and No. RS-2025-00556024 (HY), and KENTECH Research Grant No. KRG2021-01-007 (BK).
\end{acknowledgments}

\appendix
\setcounter{figure}{0}
\setcounter{table}{0}
\renewcommand{\thefigure}{A\arabic{figure}}
\renewcommand{\thetable}{A\arabic{table}}
\renewcommand{\theHfigure}{appendix.\arabic{figure}}
\renewcommand{\theHtable}{appendix.\arabic{table}}

\begin{figure}
\centering
\includegraphics[width=0.667\linewidth]{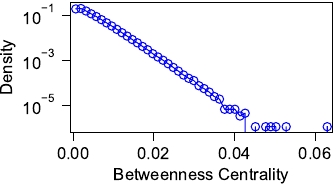}
\caption{Average density of node betweenness centrality over an ensemble of 1000 2-core pruned ER networks used in main text.}
\label{fig:BC-dist}
\end{figure}

\section{Experimental details} \label{app:details}

The underlying ER random networks are generated using \texttt{Graphs.jl} \cite{Graphs2021} for the Julia programming language \cite{Julia-2017}, with the number of nodes taken as $N_0=1000$ and the mean degree $\langle k \rangle=4$. The distribution of node betweenness centrality, obtained using the ensemble average over $1000$ configurations, is presented in Fig.~\ref{fig:BC-dist}.

When $p(t)$ is  non-integer, we implement stochastic rounding: $\lfloor p(t)\rfloor$ vehicles are created with probability $\lceil p(t) \rceil-p(t)$, and $\lceil p(t)\rceil$ vehicles are created with probability $p(t) - \lfloor p(t)\rfloor$, ensuring that the expected number of created vehicles equals $p(t)$.

The transition point $p_c$ is determined as the point beyond which the order parameter is larger than $\rho_\text{th}$. We use $\rho_\text{th}=0.002$.

For Fig.~\ref{fig:constant-p_0-qi_ai}, results are averaged over 1000 independent realizations of vehicle dynamics on a single network topology.
The $\Delta a_i$ is measured in the time interval $[t_0, t_1]$.
\begin{equation}
    \Delta a_i = \frac{1}{t_\text{obs}} \sum_{t=t_0}^{t_1-\tau} \frac{[f_i(t+\tau)-f_i(t)] + [w_i (t+\tau) - w_i(t)]}{\tau}.
\end{equation}
Similarly, $q_i$ is measured in the time interval $[t_0, t_1]$.

\section{Robustness to topology and traffic-awareness} \label{app:settings}

\begin{figure}
    \centering
    \includegraphics[width=\linewidth]{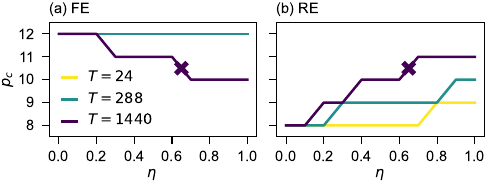}
    \caption{
        Transition point $p_c$ as a function of the fluctuation amplitude $\eta$ in BA networks under the (a) FE and (b) RE rules.
        The cross marker indicates the value of $\eta$ at which $p_c^\text{FE}(\eta) = p_c^\text{RE}(\eta)$.
        Among the tested driving periods, this crossing appears only for $T=1440$.
    }
    \label{fig:appendix-BA}
\end{figure}

\begin{figure}
    \centering
    \includegraphics[width=\linewidth]{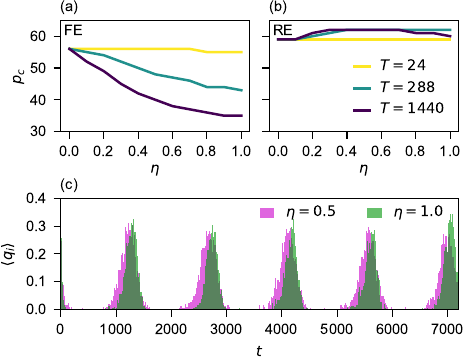}
    \caption{
        Transition point $p_c$ as a function of the fluctuation amplitude $\eta$ at $h=0.5$ under the (a) FE and (b) RE rules.
        Over the plotted parameter range, $p_c^\text{RE} > p_c^\text{FE}$.
        (c) Time evolution of the average probability $\left\langle q_i \right\rangle$ that the flowing queue is empty under the RE rule, measured at $h=0.5$ and $T=1440$.
        The average is taken over the twenty nodes with the largest BC using time windows of $20$ steps.
        The magenta and green curves correspond to $\eta=0.5$ and $\eta=1.0$, respectively.
    }
    \label{fig:appendix-high-awareness}
\end{figure}

We first examine the effect of network topology using the Barab\'asi--Albert (BA) scale-free network in Fig.~\ref{fig:appendix-BA}.
The BA networks are generated with size $N_0=1000$ and mean degree $\langle k \rangle = 4$; 2-core pruning does not remove any nodes, so $N=N_0=1000$.
The qualitative contrast between the FE and RE protocols remains robust: as the amplitude of fluctuations $\eta$ or the driving period $T$ increases, the FE protocol becomes more fragile, whereas the RE protocol can delay the onset of congestion to a higher transition point $p_c$.
However, the underlying physical mechanism differs from that in the ER network.
Because traffic is strongly concentrated around hubs and their waiting queues congest more quickly, the transition occurs at a much smaller $p_c$ than in the ER network.
Furthermore, stronger fluctuations are required to create sufficiently long intervals that allow these hubs to clear their flowing queues.
This explains why the crossover point, where the transition points $p_c$ of the RE and FE protocols are equal, occurs at higher fluctuations compared to the ER graph.

We next study a different traffic-awareness parameter using the same ER random graph as in the main text.
At $h=0.5$, the routing efficiency decreases because the hub-avoiding effect becomes excessively strong in the flowing queues.
Such degradation under overly traffic-aware routing has also been reported in previous work~\cite{echenique2005}.
Accordingly, both FE and RE exhibit lower transition points at $h=0.5$ than at $h=0.8$, as shown in Fig.~\ref{fig:appendix-high-awareness}(a) and (b).
Nevertheless, for $h=0.5$, the RE protocol consistently yields a higher $p_c$ than the FE protocol over the entire range of $\eta$.
At $T=1440$, larger $\eta$ can also slightly reduce $p_c$.
To examine the origin of this trend, we measure the time evolution of $\left\langle q_i \right\rangle$ over high-BC nodes in Fig.~\ref{fig:appendix-high-awareness}(c).
As $\eta$ increases from $0.5$ to $1.0$, the time-averaged $\left\langle q_i \right\rangle$ decreases from 0.065 to 0.053, indicating that peak traffic is not resolved efficiently in the flowing queues.

\section{Finite-size scaling analysis}

\begin{figure}
    \centering
    \includegraphics[width=\linewidth]{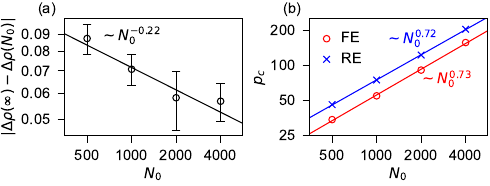}
    \caption{
        Finite-size scaling analysis of the jamming transition at $h=0.8$, $T=1440$, and $\eta=1.0$.
        (a) Scaling of the deviation of the order parameter jump $\Delta \rho$ from its asymptotic value under the FE rule.
        The solid line denotes a power-law fit, $|\Delta \rho(\infty)-\Delta \rho(N_0)| \sim N_0^{-0.22}$, with $\Delta \rho(\infty) \approx 0.3$.
        The error bars denote standard deviations among samples.
        (b) Transition point $p_c$ as a function of system size $N_0$ under the FE and RE rules.
        Symbols represent measured values, and solid lines denote power-law fits, $p_c \sim N_0^{0.73}$ for FE and $p_c \sim N_0^{0.72}$ for RE.
    }
    \label{fig:finite-size-scaling}
\end{figure}

We performed finite-size scaling analysis of the jamming transition under the FE and RE rules to examine whether the behaviors observed at $N_0=1000$ persist for different system sizes.
Four system sizes, $N_0=$ 500, 1000, 2000, and 4000, were tested for the ER 2-core network under a high-volatility setting of $T=1440$ and $\eta=1.0$ with traffic-aware routing of $h=0.8$.

We first investigated the size of the discontinuous jump in the order parameter $\Delta \rho(N_0)$ under the FE rule.
$\Delta \rho$ is measured as the separation between the two peaks of the order parameter distribution when the fraction of samples with $\rho>\rho_\text{th}$ first exceeds one half.
As shown in Fig.~\ref{fig:finite-size-scaling}(a), $|\Delta \rho (\infty)-\Delta \rho(N_0)| \sim N_0^{-0.22}$, where $\Delta \rho(\infty) \approx 0.3$.
Thus, the discontinuous transition observed in the FE protocol is not a finite-size effect, but persists in the thermodynamic limit.

Next, Fig.~\ref{fig:finite-size-scaling}(b) shows that the transition points of both FE and RE scale similarly with system size, approximately as $p_c \sim N_0^{0.73}$.
Because the two protocols have nearly the same scaling exponent, changing the system size is not expected to reverse the relative behavior observed at $N_0=1000$.
The exponent is smaller than the naive linear expectation $p_c \sim N_0$.
This can be understood from the scaling of shortest-path load in sparse ER networks: because the average shortest-path length grows logarithmically with $N$, the average BC scales as $N\log N$ rather than linearly with $N$.
Since congestion is controlled primarily by high-BC bottleneck nodes, this increasing path load provides a qualitative explanation for the sublinear growth of $p_c$ with system size.

\section{Intermediate entry protocol} \label{app:intermediate}

\begin{figure}
    \centering
    \includegraphics[width=\linewidth]{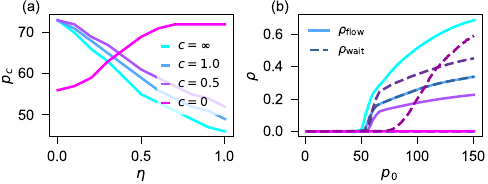}
    \caption{
    Waiting queue dependent entry protocol under $h=0.8$ and $T=1440$.
    (a) Transition point $p_c$ as a function of the fluctuation amplitude $\eta$ for different threshold coefficients $c$, including the FE limit ($c\to\infty$) and the RE limit ($c=0$).
    (b) Decomposition of the order parameter as a function of $p_0$ at $\eta=1.0$, where solid and dotted curves denote $\rho_\text{flow}$ and $\rho_\text{wait}$, respectively, and colors indicate the same $c$ values as in (a).
    }
    \label{fig:prop_th}
\end{figure}

To study an intermediate entry protocol between FE and RE, we introduce a waiting queue dependent entry rule.
In this variant, a vehicle in the waiting queue of node $i$ is allowed to enter the flowing queue when the current flowing queue length satisfies $f_i \le c w_i$, where $c$ is a threshold coefficient.
Thus, the entry condition becomes less restrictive as the waiting queue grows.
The original RE protocol is recovered at $c=0$, whereas the limit $c \to \infty$ corresponds to the FE protocol because vehicles can enter the flowing queue without restriction.

Fig.~\ref{fig:prop_th}(a) shows the transition point $p_c$ as a function of the fluctuation amplitude $\eta$ for $c=0.5$ and $c=1.0$.
For both values of $c$, $p_c$ decreases as $\eta$ increases, indicating that stronger fluctuations make the system more susceptible to jamming.
This behavior follows the FE-like trend, in contrast to the RE protocol where large fluctuations can delay the onset of congestion.

To examine these intermediate protocols more closely, Fig.~\ref{fig:prop_th}(b) decomposes the order parameter into the flowing queue contribution $\rho_\text{flow}$ and the waiting queue contribution $\rho_\text{wait}$, with $\rho=\rho_\text{flow}+\rho_\text{wait}$.
In the limiting FE and RE cases, one of these components vanishes: $\rho_\text{wait}=0$ for FE and $\rho_\text{flow}=0$ for RE.
By contrast, for $c=0.5$ and $c=1.0$, both components are positive, meaning that congestion develops in both queues.
This mixed contribution makes it difficult to isolate a single microscopic mechanism, but the macroscopic trend indicates that these intermediate protocols behave more like FE than RE under increasing fluctuations.

\bibliography{ref}

@article{chen2012,
  title = {Traffic {{Dynamics}} on {{Complex Networks}}: {{A Survey}}},
  shorttitle = {Traffic {{Dynamics}} on {{Complex Networks}}},
  author = {Chen, Shengyong and Huang, Wei and Cattani, Carlo and Altieri, Giuseppe},
  editor = {Aldashev, Gani},
  year = 2012,
  month = jan,
  journal = {Math. Probl. Eng.},
  volume = {2012},
  number = {1},
  pages = {732698},
  issn = {1024-123X, 1563-5147},
  doi = {10.1155/2012/732698},
  urldate = {2025-04-10},
  langid = {english}
}

@article{danila2006,
  title = {Optimal Transport on Complex Networks},
  author = {Danila, Bogdan and Yu, Yong and Marsh, John A. and Bassler, Kevin E.},
  year = 2006,
  month = oct,
  journal = {Phys. Rev. E},
  volume = {74},
  number = {4},
  pages = {046106},
  issn = {1539-3755, 1550-2376},
  doi = {10.1103/PhysRevE.74.046106},
  urldate = {2025-06-03},
  copyright = {http://link.aps.org/licenses/aps-default-license},
  langid = {english}
}

@article{demenezes2004,
  title = {Fluctuations in {{Network Dynamics}}},
  author = {{de Menezes}, M. Argollo and Barab{\'a}si, A.-L.},
  year = 2004,
  month = jan,
  journal = {Phys. Rev. Lett.},
  volume = {92},
  number = {2},
  pages = {028701},
  publisher = {American Physical Society},
  doi = {10.1103/PhysRevLett.92.028701},
  urldate = {2025-12-26}
}

@article{echenique2005,
  title = {Dynamics of Jamming Transitions in Complex Networks},
  author = {Echenique, P. and {G{\'o}mez-Garde{\~n}es}, J. and Moreno, Y.},
  year = 2005,
  month = jun,
  journal = {EPL},
  volume = {71},
  number = {2},
  pages = {325},
  publisher = {IOP Publishing},
  issn = {0295-5075},
  doi = {10.1209/epl/i2005-10080-8},
  urldate = {2025-04-08},
  langid = {english}
}

@inbook{erdos2011,
  title = {On the Evolution of Random Graphs},
  booktitle = {The {{Structure}} and {{Dynamics}} of {{Networks}}},
  author = {Erd{\H{o}}s, P. and R{\'e}nyi, A.},
  year = 2011,
  month = dec,
  pages = {38--82},
  publisher = {Princeton University Press},
  doi = {10.1515/9781400841356.38},
  urldate = {2025-12-30},
  collaborator = {Newman, Mark and Barab{\'a}si, Albert-L{\'a}szl{\'o} and Watts, Duncan J.},
  isbn = {978-1-4008-4135-6}
}

@article{freeman1977,
  title = {A {{Set}} of {{Measures}} of {{Centrality Based}} on {{Betweenness}}},
  author = {Freeman, Linton C.},
  year = 1977,
  journal = {Sociometry},
  volume = {40},
  number = {1},
  eprint = {3033543},
  eprinttype = {jstor},
  pages = {35--41},
  publisher = {[American Sociological Association, Sage Publications, Inc.]},
  issn = {0038-0431},
  doi = {10.2307/3033543},
  urldate = {2025-12-30}
}

@article{ohira1998,
  title = {Phase Transition in a Computer Network Traffic Model},
  author = {Ohira, Toru and Sawatari, Ryusuke},
  year = 1998,
  month = jul,
  journal = {Phy. Rev. E},
  volume = {58},
  number = {1},
  pages = {193--195},
  publisher = {American Physical Society},
  doi = {10.1103/PhysRevE.58.193},
  urldate = {2025-04-18}
}

@article{seidman1983a,
  title = {Network Structure and Minimum Degree},
  author = {Seidman, Stephen B.},
  year = 1983,
  month = sep,
  journal = {Soc. Networks},
  volume = {5},
  number = {3},
  pages = {269--287},
  issn = {03788733},
  doi = {10.1016/0378-8733(83)90028-X},
  urldate = {2025-12-30},
  copyright = {https://www.elsevier.com/tdm/userlicense/1.0/},
  langid = {english}
}

@article{takayasu2000,
  title = {Dynamic Phase Transition Observed in the {{Internet}} Traffic Flow},
  author = {Takayasu, Misako and Takayasu, Hideki and Fukuda, Kensuke},
  year = 2000,
  month = mar,
  journal = {Physica A},
  volume = {277},
  number = {1},
  pages = {248--255},
  issn = {0378-4371},
  doi = {10.1016/S0378-4371(99)00499-9},
  urldate = {2025-12-26}
}

@article{tang2009,
  title = {Self-Adjusting Routing Schemes for Time-Varying Traffic in Scale-Free Networks},
  author = {Tang, Ming and Liu, Zonghua and Liang, Xiaoming and Hui, P. M.},
  year = 2009,
  month = aug,
  journal = {Phy. Rev. E},
  volume = {80},
  number = {2},
  pages = {026114},
  publisher = {American Physical Society},
  doi = {10.1103/PhysRevE.80.026114},
  urldate = {2025-04-08}
}

@article{yan2006,
  title = {Efficient Routing on Complex Networks},
  author = {Yan, Gang and Zhou, Tao and Hu, Bo and Fu, Zhong-Qian and Wang, Bing-Hong},
  year = 2006,
  month = apr,
  journal = {Phy. Rev. E},
  volume = {73},
  number = {4},
  pages = {046108},
  issn = {1539-3755, 1550-2376},
  doi = {10.1103/PhysRevE.73.046108},
  urldate = {2025-06-03},
  copyright = {http://link.aps.org/licenses/aps-default-license},
  langid = {english}
}

@article{zhang2007a,
  title = {An Adaptive Routing Strategy for Packet Delivery in Complex Networks},
  author = {Zhang, Huan and Liu, Zonghua and Tang, Ming and Hui, P. M.},
  year = 2007,
  month = apr,
  journal = {Phys. Lett. A},
  volume = {364},
  number = {3-4},
  pages = {177--182},
  publisher = {North-Holland},
  issn = {0375-9601},
  doi = {10.1016/j.physleta.2006.12.009},
  urldate = {2025-12-30},
  langid = {american}
}

@article{liu2007,
  title = {Method to Enhance Traffic Capacity for Scale-Free Networks},
  author = {Liu, Zhe and Hu, Mao-Bin and Jiang, Rui and Wang, Wen-Xu and Wu, Qing-Song},
  year = 2007,
  month = sep,
  journal = {Phys. Rev. E},
  volume = {76},
  number = {3},
  pages = {037101},
  publisher = {American Physical Society},
  doi = {10.1103/PhysRevE.76.037101},
  urldate = {2026-01-03}
}

@article{huang2010,
  title = {Effective Strategy of Adding Nodes and Links for Maximizing the Traffic Capacity of Scale-Free Network},
  author = {Huang, Wei and Chow, Tommy W. S.},
  year = 2010,
  month = sep,
  journal = {Chaos},
  volume = {20},
  number = {3},
  pages = {033123},
  issn = {1054-1500},
  doi = {10.1063/1.3490745},
  urldate = {2026-01-03}
}

@article{echenique2004,
    title = {Improved Routing Strategies for {Internet} Traffic Delivery},
    author = {Echenique, Pablo and Gómez-Gardeñes, Jesús and Moreno, Yamir},
    journal = {Phys. Rev. E},
    volume = {70},
    number = {5},
    pages = {056105},
    year = {2004},
    doi = {10.1103/PhysRevE.70.056105},
    url = {https://link.aps.org/doi/10.1103/PhysRevE.70.056105}
}

@article{pastor-satorras2015,
  title = {Epidemic Processes in Complex Networks},
  author = {{Pastor-Satorras}, Romualdo and Castellano, Claudio and Van Mieghem, Piet and Vespignani, Alessandro},
  year = 2015,
  month = aug,
  journal = {Rev. Mod. Phys.},
  volume = {87},
  number = {3},
  pages = {925--979},
  publisher = {American Physical Society},
  doi = {10.1103/RevModPhys.87.925},
  urldate = {2026-01-28}
}

@article{moreno2004,
  title = {Dynamics of Rumor Spreading in Complex Networks},
  author = {Moreno, Yamir and Nekovee, Maziar and Pacheco, Amalio F.},
  year = 2004,
  month = jun,
  journal = {Phy. Rev. E},
  volume = {69},
  number = {6},
  pages = {066130},
  publisher = {American Physical Society},
  doi = {10.1103/PhysRevE.69.066130},
  urldate = {2026-01-28}
}

@article{jeong2000,
  title = {The Large-Scale Organization of Metabolic Networks},
  author = {Jeong, H. and Tombor, B. and Albert, R. and Oltvai, Z. N. and Barab{\'a}si, A.-L.},
  year = 2000,
  month = oct,
  journal = {Nature},
  volume = {407},
  number = {6804},
  pages = {651--654},
  publisher = {Nature Publishing Group},
  issn = {1476-4687},
  doi = {10.1038/35036627},
  urldate = {2026-01-28},
  copyright = {2000 Macmillan Magazines Ltd.},
  langid = {english}
}

@article{watts1998a,
  title = {Collective Dynamics of `Small-World' Networks},
  author = {Watts, Duncan J. and Strogatz, Steven H.},
  year = 1998,
  month = jun,
  journal = {Nature},
  volume = {393},
  number = {6684},
  pages = {440--442},
  publisher = {Nature Publishing Group},
  issn = {1476-4687},
  doi = {10.1038/30918},
  urldate = {2026-01-28},
  copyright = {1998 Macmillan Magazines Ltd.},
  langid = {english}
}

@article{faloutsos1999a,
  title = {On Power-Law Relationships of the {{Internet}} Topology},
  author = {Faloutsos, Michalis and Faloutsos, Petros and Faloutsos, Christos},
  year = {8월 30, 1999},
  journal = {SIGCOMM Comput. Commun. Rev.},
  volume = {29},
  number = {4},
  pages = {251--262},
  issn = {0146-4833},
  doi = {10.1145/316194.316229},
  urldate = {2026-01-28}
}

@article{gastner2006,
  title = {The Spatial Structure of Networks},
  author = {Gastner, M. T. and Newman, M. E.J.},
  year = 2006,
  month = jan,
  journal = {Eur. Phys. J. B},
  volume = {49},
  number = {2},
  pages = {247--252},
  issn = {1434-6036},
  doi = {10.1140/epjb/e2006-00046-8},
  urldate = {2026-01-28},
  langid = {english}
}

@article{zhang2007b,
  title = {Enhancing the Transmission Efficiency by Edge Deletion in Scale-Free Networks},
  author = {Zhang, Guo-Qing and Wang, Di and Li, Guo-Jie},
  year = 2007,
  month = jul,
  journal = {Phy. Rev. E},
  volume = {76},
  number = {1},
  pages = {017101},
  publisher = {American Physical Society},
  doi = {10.1103/PhysRevE.76.017101},
  urldate = {2026-01-28}
}

@article{huang2010a,
  title = {An Efficient Strategy for Enhancing Traffic Capacity by Removing Links in Scale-Free Networks},
  author = {Huang, Wei and Chow, Tommy W S},
  year = 2010,
  month = jan,
  journal = {J. Stat. Mech.},
  volume = {2010},
  number = {01},
  pages = {P01016},
  issn = {1742-5468},
  doi = {10.1088/1742-5468/2010/01/P01016},
  urldate = {2026-01-28},
  langid = {english}
}

@article{meloni2008,
  title = {Scaling {{Breakdown}} in {{Flow Fluctuations}} on {{Complex Networks}}},
  author = {Meloni, Sandro and {G{\'o}mez-Garde{\~n}es}, Jes{\'u}s and Latora, Vito and Moreno, Yamir},
  year = 2008,
  month = may,
  journal = {Phys. Rev. Lett.},
  volume = {100},
  number = {20},
  pages = {208701},
  publisher = {American Physical Society},
  doi = {10.1103/PhysRevLett.100.208701},
  urldate = {2026-01-28}
}

@misc{Graphs2021,
  author       = {Fairbanks, James and Besan{\c{c}}on, Mathieu and Simon, Sch{\"o}lly and Hoffiman, J{\'u}lio and Eubank, Nick and Karpinski, Stefan},
  title        = {JuliaGraphs/Graphs.jl: an optimized graphs package for the Julia programming language},
  year         = 2021,
  url = {https://github.com/JuliaGraphs/Graphs.jl/}
}

@article{Julia-2017,
    title={Julia: A fresh approach to numerical computing},
    author={Bezanson, Jeff and Edelman, Alan and Karpinski, Stefan and Shah, Viral B},
    journal={SIAM Rev.},
    volume={59},
    number={1},
    pages={65--98},
    year={2017},
    publisher={SIAM},
    doi={10.1137/141000671},
    url={https://epubs.siam.org/doi/10.1137/141000671}
}

@article{ling2010,
  title = {Global Dynamic Routing for Scale-Free Networks},
  author = {Ling, Xiang and Hu, Mao-Bin and Jiang, Rui and Wu, Qing-Song},
  year = 2010,
  month = jan,
  journal = {Phys. Rev. E},
  volume = {81},
  number = {1},
  pages = {016113},
  publisher = {American Physical Society},
  doi = {10.1103/PhysRevE.81.016113},
  urldate = {2026-01-29}
}

@article{danila2007,
  title = {Transport Optimization on Complex Networks},
  author = {Danila, Bogdan and Yu, Yong and Marsh, John A. and Bassler, Kevin E.},
  year = 2007,
  month = jun,
  journal = {Chaos},
  volume = {17},
  number = {2},
  pages = {026102},
  issn = {1054-1500},
  doi = {10.1063/1.2731718},
  urldate = {2026-01-29}
}

@article{kaitian2019,
  title = {Routing Optimization Considering Real-Time Traffic Condition and Travel Distance},
  author = {Kaitian, Luo and Gang, Liu},
  year = 2019,
  month = jun,
  journal = {Int. J. Mod. Phys. B},
  volume = {33},
  number = {16},
  pages = {1950169},
  publisher = {World Scientific Publishing Co.},
  issn = {0217-9792},
  doi = {10.1142/S0217979219501698},
  urldate = {2026-01-29}
}

@article{papageorgiou2002,
  title = {Freeway Ramp Metering: An Overview},
  shorttitle = {Freeway Ramp Metering},
  author = {Papageorgiou, M. and Kotsialos, A.},
  year = 2002,
  month = dec,
  journal = {IEEE Trans. Intell. Transp. Syst},
  volume = {3},
  number = {4},
  pages = {271--281},
  issn = {1524-9050},
  doi = {10.1109/TITS.2002.806803},
  urldate = {2026-07-14},
  copyright = {https://ieeexplore.ieee.org/Xplorehelp/downloads/license-information/IEEE.html},
  langid = {english}
}

@article{papageorgiou,
  title = {ALINEA: A local feedback control law for on-ramp metering},
  author = {Papageorgiou, Markos and Haj-Salem, Habib and Blosseville, Jean-Marc},
  langid = {english},
  journal={Transp. Res. Rec.},
  volume = {1320},
  number = {1},
  year={1991},
  url={https://trid.trb.org/View/365587}
}

\end{document}